%% file: main.tex
\documentclass[conference,a4paper]{IEEEtran}
\usepackage{cite}
\usepackage{amsmath,amssymb,amsfonts,bm}
\usepackage{mathrsfs}
\usepackage[linesnumbered,ruled,vlined]{algorithm2e}
\usepackage{amsthm}
\usepackage{dsfont}
\usepackage{algorithmic}
\usepackage{subfigure}
\usepackage{graphicx}
\usepackage{textcomp}
 \usepackage{threeparttable}
\usepackage{xcolor}
\usepackage{bbm}
\usepackage[nolist,withpage]{acronym}
\usepackage{array}
\usepackage{tikz}
\usepackage{pgfplots}
\usepackage{dsfont}
\pgfplotsset{compat=newest}

\definecolor{copperrose}{rgb}{0.6, 0.4, 0.4}
\definecolor{azure}{rgb}{0.0, 0.5, 1.0}
\definecolor{ashgrey}{rgb}{0.7, 0.75, 0.71}
\definecolor{chestnut}{rgb}{0.8, 0.36, 0.36}
\definecolor{airforceblue}{rgb}{0.36, 0.54, 0.66}
\definecolor{cadmiumorange}{rgb}{0.93, 0.53, 0.18}
\definecolor{bleudefrance}{rgb}{0.19, 0.55, 0.91}
\definecolor{carolinablue}{rgb}{0.6, 0.73, 0.89}
\definecolor{blue(ncs)}{rgb}{0.0, 0.53, 0.74}
\definecolor{dodgerblue}{rgb}{0.12, 0.56, 1.0}
\definecolor{cssgreen}{rgb}{0.0, 0.5, 0.0}
\definecolor{cadmiumgreen}{rgb}{0.0, 0.42, 0.24}
\definecolor{cadmiumorange}{rgb}{0.93, 0.53, 0.18}
\definecolor{amaranth}{rgb}{0.9, 0.17, 0.31}
\definecolor{bluegray}{rgb}{0.4, 0.6, 0.8}
\definecolor{cerulean}{rgb}{0.0, 0.48, 0.65}
\definecolor{ceil}{rgb}{0.57, 0.63, 0.81}

\usepackage{etoolbox}
\makeatletter
\newif\if@in@acrolist
\AtBeginEnvironment{acronym}{\@in@acrolisttrue}
\newrobustcmd{\LU}[2]{\if@in@acrolist#1\else#2\fi}

\newcommand{\ACF}[1]{{\@in@acrolisttrue\acf{#1}}}

\begin{document}

\input{acronyms.tex}

\title{Joint Bit-Partitioning and Modulation \\ Design for Digital AirComp
}

\author{Xiaojing Yan, Carlo Fischione\\
	\normalsize School of Electrical Engineering and Computer Science, KTH Royal Institute of Technology, Stockholm, Sweden\\
	\normalsize Email: \{xiay, carlofi\}@kth.se
}

\newtheorem{theorem}{Theorem}
\newtheorem{prop}{Proposition}
\newtheorem{lem}{Lemma}
\newtheorem{rem}{Remark}

\maketitle
\thispagestyle{empty}
\pagestyle{empty}

\begin{abstract}

For digital over-the-air computation, the ChannelComp framework has recently been proposed to design digital modulations to compute any arbitrary function over a multiple access channel. To reduce modulation design complexity while increasing  computation reliability, this paper integrates a bit-partitioning procedure into ChannelComp. The key process is to partition the input bit sequence into several groups, map each group to a single modulation symbol and transmit the encoded symbol sequence across multiple time slots. With the objective to maximize a worst-case constellation distance, we develop two bit-partitioning methods. In uniform bit-partitioning, bits are evenly distributed across groups and modulation is designed via a max–min optimization, which is handled by a \ac{CCCP} that solves a sequence of second-order
cone programming subproblems. In \ac{IABP}, the bit allocation is adapted to the significance of individual bit positions, and the modulation and partitioning are jointly optimized. To keep the overall complexity manageable, 
simulated annealing is employed in the outer loop to update the partitioning, while a \ac{CCCP}-based solver is used in the inner loop for modulation design. Numerical results show that both methods provide robust computation in noisy channels, and IABP achieves up to a 5 dB reduction in computation error compared to Sequential Modulation for AirComp, especially for product computation.


\end{abstract}


\section{Introduction}

\acresetall

The rapid growth of wireless devices and new services in 6G is driving an explosion of data from intelligent \ac{IoT} devices. A large amount of data needs to be gathered and delivered to servers for computation, analysis, and decisions~\cite{liu2020over}. In response, \ac{AirComp} offers a task-oriented alternative in which devices transmit simultaneously over a \ac{MAC}, allowing the receiver to directly estimate a desired function output of the distributed data. By leveraging waveform superposition with suitable function representations, AirComp provides low-latency, spectrum-efficient aggregation aligned with the computation goal~\cite{csahin2023survey}. This is especially attractive for distributed learning, where gradient and model-update aggregation can be done over the air, improving the usage of compute resources for large-scale training systems~\cite{amiri2020federated}.

Early studies in \ac{AirComp} focused on computing linear functions over the \ac{MAC} and were later extended to nomographic functions. These approaches, however, relied on analog amplitude modulation, which is difficult to integrate with standard wireless hardware and  sensitive to noise and fading~\cite{goldenbaum2013harnessing}. This has led to a move toward digital \ac{AirComp}, where simple constellations such as BPSK and FSK have been used for learning tasks over wireless networks~\cite{bernstein2018signsgd,csahin2021distributed}. These digital schemes are limited to specific function computations and often result in inefficient resource utilization~\cite{perez2024waveforms}. To overcome these challenges, ChannelComp has been proposed as a general modulation-based framework for arbitrary function computation, offering reliable computation~\cite{saeed2023ChannelComp}.

However, in ChannelComp, each function input is encoded into a single digital modulation symbol, whereas recent multi-symbol designs advance beyond this by exploiting transmission across multiple time slots.
In~\cite{yan2025remac}, \ac{ReMAC} extends ChannelComp to enable selective symbol repetition and improve computation accuracy, but its \ac{SDR}-based method scales poorly in large-scale networks. As a lower complexity framework, Bit-Slicing~\cite{liu2025digital} allocates bit sequences into slices and maps each sliced integer to square quadrature amplitude modulation symbols for transmission, yet it is mainly effective for the sum function. To enhance the flexibility of modulation design, \ac{SeMAC} encodes each input into a sequence of modulation symbols and employs an iterative algorithm based on successive convex approximations to reduce computational cost while retaining high computation accuracy~\cite{yan2025multi}. However, it does not consider bit-level significance, since the erroneous reception of more significant bits leads to larger computational error. 

In this paper, we integrate the ChannelComp framework with a bit-partitioning procedure for digital \ac{AirComp}. The idea is to partition the input bit sequence into several groups, encode each bit group to a single modulation symbol, and transmit the resulting symbol sequence over multiple time slots.
To guarantee a robust computation, we develop two bit-partitioning methods. The first, \ac{UBP}, assigns an equal number of bits to each group. For this case, we formulate the modulation design as a max–min optimization that maximizes the worst-case constellation distance. Although the problem is non-convex, it can be handled by a \ac{CCCP} that iteratively solves a sequence of relaxed \ac{SOCP} subproblems. The second method, \ac{IABP}, allocates bits according to their positional significance and jointly optimizes both the partitioning and modulation under the same max–min criterion. To keep the complexity tractable, we place the \ac{CCCP}-based modulation design inside an outer \ac{SA} loop that updates the partitioning of input bit sequence.
Numerical results show that both methods achieve robust computation, and \ac{IABP} reduces computation error by up to 5~dB compared to \ac{SeMAC}, particularly for product function.


\section{System Model}\label{sec:Systemmodel}

\input{Figures/Fig-system}

In this section, we present the multi-symbol digital \ac{AirComp} framework integrated with bit-partitioning and outline its basic operations. The communication process involves three main blocks: encoding, transmission and decoding.

\subsection{Encoding}  \label{subsec:signal_model}

Consider a network with $K$ single-antenna nodes transmitting data to a \ac{CP} over a \ac{MAC}. The goal of the \ac{CP} is to compute a function $f(x_1,\ldots,x_K):\mathbb{R}^K \mapsto \mathbb{R}$, where $x_k \in \mathbb{R}$ represents the input value held by node $k$. Before transmission, the encoding procedure $\mathscr{E}_k(\cdot):\mathbb{R} \mapsto \mathbb{C}^L$ of node $k$ is elaborated as follows.

\subsubsection{Quantization}

Each node $k$ first quantizes its input value $x_k$ into a finite set $\mathcal{X}_f$ of $Q=2^B$ quantization levels using a $B$-bit quantizer $\mathcal{Q}_B(\cdot):\mathbb{R} \mapsto \mathcal{X}_f$. 
The quantized value is denoted by $\tilde{x}_k := \mathcal{Q}_B(x_k) \in \mathcal{X}_f$, and can be represented by a binary vector $\bm{d}_k = [d_{k,1}, \dots, d_{k,B}] \in \{0,1\}^B$, where $d_{k,1}$ is the most significant bit and $d_{k,B}$ is the least significant bit. 

\subsubsection{Bit-Partitioning}

For each node $k$, the $B$-bit binary sequence $\bm{d}_k$ is partitioned into $L$ groups based on a group-size vector $\bm{b} = [b_1, \dots, b_L] \in \mathbb{Z}_+^L$, where $b_\ell$ indicates the number of bits in group $\ell$ and $\sum_{\ell=1}^L b_\ell = B$. Let $s_\ell = \sum_{j=1}^{\ell} b_j$ denote the cumulated bit width of the first $\ell$ groups, with $s_0 = 0$. Then, the bit subsequence assigned to group $\ell$ can be defined as $\bm{d}_{k,\ell} = [d_{k,s_{\ell-1}+1}, \dots, d_{k,s_\ell}] \in \{0,1\}^{b_\ell}$, and its corresponding integer value $\tilde{x}_{k,\ell}$ is computed as:
\begin{equation}
\tilde{x}_{k,\ell} = \sum\nolimits_{j=1}^{b_\ell} d_{k,s_{\ell-1}+j} \cdot 2^{b_\ell - j}, \quad \text{for } \ell \in [L]. 
\end{equation}
This procedure, referred to as \emph{bit-partitioning}, decomposes the quantized input $\tilde{x}_k$ into $L$ grouped integers via a Spliter $\mathcal{S}_k(\cdot):\mathbb{R}\mapsto \mathbb{N}^L$, i.e., $[\tilde{x}_{k,1}, \dots, \tilde{x}_{k,L}]:=\mathcal{S}_k(\tilde{x}_k)$, where each element lies in the range $\{0, \dots, 2^{b_\ell} - 1\}$.

\subsubsection{Digital Modulation}

Each grouped integer $\tilde{x}_{k,\ell}$ is modulated into a complex modulation symbol $\vec{x}_{k,\ell} \in \mathbb{C}$. The complete modulation sequence is generated by an encoder $\mathcal{E}_k(\cdot):\mathbb{N}^L \mapsto \mathbb{C}^L$, i.e., $[\vec{x}_{k,1}, \ldots, \vec{x}_{k,L}]:=\mathcal{E}_k(\tilde{x}_{k,1},\ldots,\tilde{x}_{k,L})$, with each symbol preparesd to transmit in a distinct time slot. 
In this way, the input value $x_k$ is encoded into a sequence of $L$ modulated symbols, i.e., $[\vec{x}_{k,1},\ldots,\vec{x}_{k,L}]=\mathscr{E}_k(x_k)$.

\subsection{Transmission}
All $K$ nodes simultaneously transmit their encoded modulated symbol sequences over the \ac{MAC}.
Assuming perfect synchronization\footnote{{\color{black}Note that perfect synchronization is a common assumption in \ac{AirComp} literature~\cite{perez2024waveforms}. In practical systems, techniques such as frame timing and carrier frequency
offset estimation can be applied for achieving synchronization~\cite{guo2021over}.}}, the CP receives the superimposed signals across $L$ time slots as follows:
\begin{equation}
\vec{y}_\ell = \sum\nolimits_{k=1}^{K} h_{k,\ell} p_{k,\ell} \vec{x}_{k,\ell} + \vec{z}_\ell, \quad \forall \ell \in [L],  
\end{equation}
where $h_{k,\ell}$ is the channel coefficient, $p_{k,\ell} $ is the transmission power, and $\vec{z}_\ell  \sim \mathcal{N}(0, \sigma_z^2)$ is the additive white Gaussian noise. To compensate for the fading effects, optimal power control is applied by inverting the channel~\cite{cao2019optimal}, where the power is set as $p_{k,\ell} = h_{k,\ell}^*/|h_{k,\ell}|^2$. Consequently, the received signal at the \ac{CP} in each time slot simplifies to:
\begin{equation}
\label{eq:nofading}
\vec{y}_\ell = \sum\nolimits_{k=1}^{K} \vec{x}_{k,\ell} + \vec{z}_\ell, \quad \forall \ell \in [L].
\end{equation}
Collecting all the received symbols, we define the received sequence as $\bm{y}:=[\vec{y}_1,\ldots,\vec{y}_L] \in \mathbb{C}^L$.

\subsection{Decoding}

In the \ac{CP}, the decoding process $\mathscr{D}(\cdot):\mathbb{C}^L \mapsto \mathbb{R}$ consists of two main components: symbol estimation and tabular mapping.

\subsubsection{Symbol Estimation}

As discussed in~\cite{liu2025digital}, the optimal estimator at the \ac{CP} can be derived using the maximum a posteriori criterion. However, following similar steps in~\cite{saeed2023ChannelComp}, we adopt a maximum likelihood estimator $\mathcal{M}(\cdot):\mathbb{C}^L \mapsto \mathbb{C}^L$. In each time slot $\ell$, the \ac{CP} estimates the aggregated constellation point
$\vec{v}_\ell := \sum\nolimits_{k=1}^{K} \vec{x}_{k,\ell}$ from the noisy observation $\vec{y}_\ell$, and the full estimated sequence $\bm{v}:=[\vec{v}_1,\ldots,\vec{v}_L] \in \mathbb{C}^L$ is recovered via $\bm{v}=\mathcal{M}(\bm{y})$.

\subsubsection{Tabular Mapping}

After recovering the estimated sequence $\bm{v}$, the \ac{CP} applies a $B$-bit tabular operator $\mathcal{T}_B(\cdot): \mathbb{C}^L \mapsto \mathcal{Y}_f$ that matches the input quantization precision. The final estimate is $\hat{f} = \mathcal{T}_B(\bm{v}) \in {\mathcal{Y}}_f$, where ${\mathcal{Y}}_f$ denotes the set of quantized output values of function $f$. Therefore, the estimated output is recovered from the received symbol sequence, i.e., $\hat{f}=\mathscr{D}(\bm{y})$.

\section{Problem Statement}

This section introduces the priciples for designing the encoder and decoder under bit-partitioning to enable robust computation.
Since we partition the $B$ input bits into groups based on $\bm{b}$, for each group $\ell$, let $Q_\ell = 2^{b_\ell}$ denote the number of possible integer values that can be represented by the  $b_\ell$ bits. Then, for node $k$, the $Q_{\ell}$ integer values are characterized by a codebook of  complex-valued symbols, denoted by $\bm{x}_{k,\ell} := [\vec{x}_{k,\ell}^{(1)}, \ldots, \vec{x}_{k,\ell}^{(Q_\ell)}]^\mathsf{T} \in \mathbb{C}^{Q_\ell}$.
By stacking the codebooks from all $K$ nodes, the modulation vector for group $\ell$ can be defined as $\bm{x}_\ell := [\bm{x}_{1,\ell}^\mathsf{T}, \ldots, \bm{x}_{K,\ell}^\mathsf{T}]^\mathsf{T} \in \mathbb{C}^{N_\ell}$, where $N_\ell=KQ_\ell$. Additionally, we impose a unit-norm constraint on each modulation vector and denote the feasible set by $\mathcal{X}:=\{\|\bm{x}_\ell\|_2 \leq1, \forall \ell \in [L]\}$.

Now consider a noiseless \ac{MAC} and two input sets 
$x_1^{(i)}, \dots, x_K^{(i)}$ and $x_1^{(j)}, \dots, x_K^{(j)}$, 
which yield distinct function values $f^{(i)} \neq f^{(j)}$. 
Since each input value is modulated into a sequence of $L$ complex symbols, the resulting aggregated constellation sequences 
$\bm{v}^{(i)} = \sum_{k=1}^K \mathscr{E}_k(x_k^{(i)})$ 
and 
$\bm{v}^{(j)} = \sum_{k=1}^K \mathscr{E}_k(x_k^{(j)})$. 
For each time slot $\ell$, the constellation point associated with the $i$-th input case can be expressed as $v^{(i)}_\ell = \bm{a}_{i,\ell}^{\mathsf T}(\bm{b}) \bm{x}_\ell$. Specifically,
$\bm{a}_{i,\ell}(\bm{b})\in\{0,1\}^{N_\ell}$ is a binary vector selects the support of the $\ell$-th grouped modulation symbols associated with $f^{(i)}$, and its dimension is determined by the group-size vector $\bm{b}$.

For reliable computation, the aggregated constellation sequences $\bm{v}^{(i)}$ and $\bm{v}^{(j)}$ must be distinguishable at the \ac{CP} whenever the corresponding quantized outputs differ, leading to the following constraints: 
\begin{align}
\label{eq:CompCond_nonsmooth}
{\rm if}~\hat{f}^{(i)} \neq \hat{f}^{(j)} \Rightarrow \bm{v}^{(i)} \neq \bm{v}^{(j)}, \quad \forall (i,j) \in [M]^2,
\end{align}
where $M$ is the number of distinct quantized function outputs. 
To make this requirement quantitative, consider the squared Euclidean distance between two constellation sequences:
\begin{align}
\label{eq:Euclidean distance} \nonumber
\|\bm{v}^{(i)} - \bm{v}^{(j)}\|_2^2
= & \sum\nolimits_{\ell=1}^{L} 
\Big|(\bm{a}_{i,\ell}(\bm{b}) - \bm{a}_{j,\ell}(\bm{b}))^{\mathsf T}\bm{x}_\ell\Big|^2 \\
= & \sum\nolimits_{\ell=1}^{L} \bm{x}_\ell^{\mathsf H}\bm{D}_{i,j,\ell}(\bm{b})\,\bm{x}_\ell,
\end{align}
where $\bm{D}_{i,j,\ell}(\bm{b}):=[\bm{a}_{i,\ell}(\bm{b}) - \bm{a}_{j,\ell}(\bm{b})][\bm{a}_{i,\ell}(\bm{b}) - \bm{a}_{j,\ell}(\bm{b})]^{\mathsf{T}}$. 
A larger Euclidean distance makes two received sequences easier to distinguish at the receiver. However, this distance only reflects geometry in the signal space but ignores how far the corresponding function values are from each other. When two function outputs differ more, their signal representations should be more separated than those for outputs that are closer.
To model this effect, we scale the Euclidean distance by $\Delta \hat{f}_{i,j} = \epsilon|\hat{f}^{(i)} - \hat{f}^{(j)}|$ with $\epsilon>0$. For pairs with $\Delta \hat{f}_{i,j} \neq 0$, the scaled distance is defined as
\begin{equation}
\label{eq:scaled-distance} 
r_{i,j}(\{\bm{x}_\ell\},\bm{b}) :=
\frac{1}{\Delta \hat{f}_{i,j}}
\sum\nolimits_{\ell=1}^{L} \bm{x}_\ell^{\mathsf H}\bm{D}_{i,j,\ell}(\bm{b})\,\bm{x}_\ell,
\end{equation}
and let $\tilde{M}$ indicates the number of the pairwise distances. We then quantify the computation robustness using the worst-case distance, given by the minimum over these pairs:
\begin{equation}
\label{eq:worst-case-distance}
d_{\min}(\{\bm{x}_\ell\},\bm{b}) := 
\min_{\Delta \hat{f}_{i,j} \neq 0} r_{i,j}(\{\bm{x}_\ell\},\bm{b}).
\end{equation}
Maximizing $d_{\min}({\bm{x}_\ell},\bm{b})$ encourages that output pairs with greater differences are separated by proportionally larger constellation distances, making them less likely to be misestimated over noisy \ac{MAC}s.  
Based on this principle, we next introduce two bit-partitioning methods to achieve reliable computation.

\section{The Proposed Bit-Partitioning Methods}

In this section, we develop two specific bit-partitioning methods for multi-symbol digital \ac{AirComp}. \ac{UBP} distributes the same number of bits to each group and focuses on the modulation design that enlarges the worst-case distance for robust computation. \ac{IABP} allocates bits according to the significance of bit positions and couples bit-partitioning with modulation design, but at the cost of a larger design complexity.

\subsection{Uniform Bit-Partitioning}
\label{subsec:uniform bit-grouping}

We first focus on the uniform partitioning, where the total bits are evenly distributed across all groups, and the entire encoding and decoding flow is illustrated in~\ref{fig:uniform bit-grouping}. If $B$ is not divisible by $L$, the leftmost group is zero-padded. Specifically, each group is assigned
$b = \lceil B/L \rceil$ bits and employs the identical modulation vector $\bm{x}\in \mathbb{C}^{N}$, where $N=K\cdot2^b$. Under this setting, the worst-case distance in \eqref{eq:worst-case-distance} reduces to
\begin{align}
d_{\min}(\bm{x};\bm{b}) 
= \min_{\Delta \hat{f}_{i,j}\neq 0} 
\frac{1}{\Delta \hat{f}_{i,j}}\bm{x}^{\mathsf H}\bm{D}_{i,j}\,\bm{x}.
\end{align}
Here, since $B$ and $L$ fix $\bm{b}$, we drop parameter $\bm{b}$ from the notation and write $\bm{D}_{i,j} := \sum_{\ell=1}^L \bm{D}_{i,j,\ell}(\bm{b})$. Accordingly, to design
the constellation diagram, we pose the following optimization to maximize $d_{\min}(\bm{x};\bm{b})$ over the feasible set $\mathcal{X}$:
\begin{align}
\mathcal{P}_{0}:=
\max_{\bm{x}\in \mathcal{X}}\; d_{\min}(\bm{x};\bm{b}).
\end{align}
Introducing an auxiliary variable $c>0$ to represent the worst-case distance, then Problem $\mathcal{P}_0$ is equivalently written as:
\begin{subequations}
\label{eq:equivalent P0}
\begin{align}
\nonumber
 \mathcal{P}_{1} := & \max_{\bm{x}\in \mathcal{X}, c>0} \; c\\ 
 {\rm s.t.} 
 \quad & c\Delta \hat{f}_{i,j} - \bm{x}^{\mathsf H}\bm{D}_{i,j}\,\bm{x} \leq 0, \quad \text{for } \Delta \hat{f}_{i,j}\neq 0.
\end{align}
\end{subequations}
Problem $\mathcal{P}_1$ is a \ac{DC} programming, where the quadratic term $-\bm{x}\bm{D}_{i,j}\bm{x}$ is concave. We can address it via \ac{CCCP}~\cite{yuille2003concave}, a heuristic algorithm that iteratively majorizes the concave part and solves a sequence of convex surrogates. Let $\bm{x}^{(t)}$ be the solution at iteration $t-1$. At iteration $t$, we replace concave terms with a convex upper bound around $\bm{x}^{(t)}$, and obtain the convexified subproblem:
\begin{subequations}
\label{eq:convex subproblem P0}
\begin{align}
\nonumber
 \mathcal{P}_{2} := & \min_{\bm{x}\in \mathcal{X}, c>0} \; -c \\  \nonumber
 {\rm s.t.} 
 \quad & c\Delta \hat{f}_{i,j} - 2\Re\{\bm{x}^{(t)\mathsf{H}}\bm{D}_{i,j}\bm{x}\} +\bm{x}^{(t)\mathsf{H}}\bm{D}_{i,j}\bm{x}^{(t)}  \leq 0, \\ 
 \quad &  \text{for } \Delta \hat{f}_{i,j}\neq 0,
\end{align}
\end{subequations}
Problem $\mathcal{P}_2$ is a \ac{SOCP}, which can be handled by standard CVX solvers~\cite{alizadeh2003second}. 
We terminate the procedure when the improvement in the objective is small, i.e., $c^{(t+1)}-c^{(t)} < \delta$.
The final iterate is returned as an approximate solution to $\mathcal{P}_0$, and the complete procedure is summaried in Algorithm~\ref{Alg:uniform algorithm}.

\input{Figures/uniform_Alg}

\begin{theorem} \label{theorem:convergence}
Let $\{\bm{x}^{(t)},c^{(t)}\}$ be the sequence generated by Algorithm~\ref{Alg:uniform algorithm}, then the objective $\{-c^{(t)}\}_{t=0}^{\infty}$ is non-increasing and will converge. All the limit points of $\{\bm{x}^{(t)}\}_{t=0}^{\infty}$ are stationary points of the original problem $\mathcal{P}_0$.

\end{theorem}

\begin{proof}
See Appendix~\ref{app:convergence}.    
\end{proof}

\subsection{Importance-Adaptive Bit-Partitioning}

In the \ac{UBP}, all groups are assigned equal bit widths without accounting for the fact that estimation errors in more significant groups cause larger deviations in the computation output.
To address this limitation, we propose an adaptive method that couples the modulation design with bit-partitioning. The idea is to assign fewer bits to more significant groups so as to enlarge the minimum constellation distance, thereby lowering the probability of their symbol estimation errors. The corresponding model is illustrated in~\ref{fig:importance-aware bit-grouping}.

Concretely, we keep vector $\bm{b}$
in non-increasing order to prioritize the accurate aggregation
of more significant groups. To incorporate relative importance into the distance measure, we introduce group-wise weights employing a Gaussian profile:
\begin{equation}
\label{eq:Gaussian weight}
w_\ell = \exp\!\left(-\frac{(\ell-1)^2}{2\sigma^2}\right), \quad \forall \ell \in [L],
\end{equation}
with decay parameter $\sigma>0$, and normalize them as $\tilde{w}_\ell = w_\ell/\sum_{\ell=1}^L w_\ell$.
Then, we extend the scaled distance in~\eqref{eq:scaled-distance} as:
\begin{equation}
\label{eq:extended scaled distance} 
\tilde r_{i,j}(\{\bm{x}_\ell\},\bm{b}) 
:= \frac{1}{\Delta \hat{f}_{i,j}}\sum\nolimits_{\ell=1}^L \tilde{w}_\ell\,\bm{x}_\ell^{\mathsf H}\bm{D}_{i,j,\ell}(\bm{b})\bm{x}_\ell,
\end{equation}
and define the corresponding weighted worst-case distance as:
\begin{equation}
\label{eq:extended worst-case distance}
\tilde d_{\min}(\{\bm{x}_\ell\},\bm{b}) 
:= \min_{\Delta \hat{f}_{i,j}\neq 0} \tilde r_{i,j}(\{\bm{x}_\ell\},\bm{b}).
\end{equation}
\input{Figures/importance_aware_Alg}

\input{Figures/uniform_bit_grouping}

Consequently, to jointly design
the constellation diagram and bit sequence partitioning, we pose the following optimization
problem by maximizing $\tilde d_{\min}(\{\bm{x}_\ell\},\bm{b})$, as given by:
\begin{subequations}
\label{eq:smooth original problem}
\begin{align}
\nonumber
 \mathcal{P}_{3} := & \max_{\{\bm{x}_\ell\} \in \mathcal{X},\bm{b}} \; \tilde d_{\min}(\{\bm{x}_\ell\},\bm{b})\\ 
 {\rm s.t.} 
 \quad & \sum\nolimits_{\ell=1}^L b_{\ell}=B, \quad b_1 \leq \ldots \leq b_L, \quad b_{\ell} \in \mathbb{Z}_+.
\end{align}
\end{subequations}
Problem $\mathcal{P}_3$ is a non-convex and involves a discrete search over $\bm{b}$ coupled with continuous optimization over $\{\bm{x}_\ell\}$. We can adopt a bi-level approach, where we employ \ac{SA}~\cite{bertsimas1993simulated} to update $\bm{b}$ in the outer loop and optimize $\{\bm{x}_\ell\}$ via \ac{CCCP} in the inner loop.
The annealing procedure begins with an initial vector $\bm{b}^{(0)}$. In each iteration $r$, a candidate $\bm{b}'$ is generated by shifting one bit between adjacent groups while preserving the non-decreasing order and the total bit constraint. To evaluate a group-size vector, we introduce an energy function $\mathcal{N}(\bm{b})$, determined as the objective value of $\mathcal{P}_3$ with $\bm{b}$ fixed. This leads to the problem:
\begin{equation}
\mathcal{P}_{4}:= 
\max_{\{\bm{x}_\ell\} \in \mathcal{X}} \; \tilde d_{\min}(\{\bm{x}_\ell\};\bm{b}).
\end{equation}
Problem $\mathcal{P}_4$ can be regarded as the inner optimization over the modulation vectors $\{\bm{x}_\ell\}$. It retains the same \ac{DC} structure as $\mathcal{P}_0$ and can be relaxed to the following formulation:
\begin{subequations}
\label{eq:relaxed convex subproblem P4}
\begin{align}
\nonumber
 \mathcal{P}_{5} := & \min_{\{\bm{x}_\ell\}\in \mathcal{X}, c} \; -c \\  \nonumber
 {\rm s.t.} 
 \quad & c\Delta \hat{f}_{i,j} - 2\Re\left\{\sum\nolimits_{\ell=1}^L\tilde{w}_\ell\bm{x}_\ell^{(t)\mathsf{H}}\bm{D}_{i,j,\ell}\bm{x}_\ell\right\} \\ \nonumber
 & +\sum\nolimits_{\ell=1}^L\tilde{w}_\ell\bm{x}_\ell^{(t)\mathsf{H}}\bm{D}_{i,j,\ell}\bm{x}_\ell^{(t)}  \leq 0,  \text{for } \Delta \hat{f}_{i,j}\neq 0.
\end{align}
\end{subequations}
Iterating the convexify–solve steps in \ac{CCCP} yields $\{\hat{\bm{x}}_\ell\}$ for the current group-size vector $\bm{b}$. We then substitute $\{\hat{\bm{x}}_\ell\}$ back into $\mathcal{P}_4$ to obtain the corresponding energy value. Based on this value, the acceptance probability for moving from $\mathbf{b}^{(r)}$ to $\mathbf{b}'$ is given by:
\begin{equation}
\label{eq:annealing probability}
p = \min\left(1, \exp\left(\frac{\mathcal{N}(\bm{b}') - \mathcal{N}(\bm{b}^{(r)})}{\phi_r}\right)\right),
\end{equation}
where $\phi_t>0$ is the annealing temperature following a cooling schedule $\phi_{r+1}=\alpha\phi_r$ with $\alpha\in(0,1)$.
Accordingly, we set $\mathbf{b}^{(r+1)}=\mathbf{b}'$ with probability $p$. Otherwise, we keep $\mathbf{b}^{(r+1)}=\mathbf{b}^{(r)}$. The annealing process stops when $\phi_r<\phi_{\min}$, and a global asymptotic of this process can be achieved under the logarithmic cooling schedule~\cite{bertsimas1993simulated}. The final iterate is taken as an approximate solution to the original $\mathcal{P}_3$.
The combined \ac{SA} and \ac{CCCP} procedure is summarized in Algorithm~\ref{Alg:importance-aware algorithm}.

\subsection{Complexity Analysis}

We analyze the complexity of the proposed Algorithm~\ref{Alg:uniform algorithm} and Algorithm~\ref{Alg:importance-aware algorithm} in this subsection. In the uniform case, each itertaion of Algorithm~\ref{Alg:uniform algorithm} solves the \ac{SOCP} subproblem $\mathcal{P}_2$ via an interior-point method whose cost is $\mathcal{O}(N^3+\tilde{M}N^2)$, leading to an overall complexity $\mathcal{O}(S[N^3+\tilde{M}N^2])$ for $S$ iterations in \ac{CCCP} procedure~\cite{alizadeh2003second}. In the importance-adaptive case, under a fixed $\bm{b}$, the inner modulation subproblem costs $\mathcal{O}(S[(\sum_\ell N_\ell)^3+\tilde{M}(\sum_\ell N_\ell)^2])$. Across $T$ annealing updates of $\bm{b}$, the total cost of Algorithm~\ref{Alg:importance-aware algorithm} becomes $\mathcal{O}(TS[(\sum_\ell N_\ell)^3+\tilde{M}(\sum_\ell N_\ell)^2])$. Since $\sum_{\ell}N_\ell \geq N$ and an outer \ac{SA} loop is involved, Algorithm~\ref{Alg:importance-aware algorithm} entails higher computational complexity than Algorithm~\ref{Alg:uniform algorithm}. Moreover, due to bit-partitioning procedure, the optimization dimension $N$ or $\sum_{\ell}N_\ell$ in our \ac{SOCP} subproblems are reduced relative to the single-symbol design. To scale to large-scale networks, we can adopt an offline workflow, where the modulation and bit-partitioning co-design only needs to be solved once in advance and then reused during real-time applications.



\section{Numerical Experiments}\label{sec:Num}

In this section, we evaluate the performance of the two proposed bit-partitioning strategies, and make a comparison with the existing multi-symobl modulation schemes, i.e., SeMAC,  ReMAC, and Bit-Slicing. Specifically, the performance is evaluated using the \ac{NMSE} metric, defined as 
\begin{equation}
\text{\ac{NMSE}}:=\frac{\sum\nolimits_{j=1}^{N_s}|f^{(i)}-\hat{f}_j^{(i)}|^2}{N_s|f_{\max} - f_{\min}|^2},    
\end{equation}
where $N_s=100$ is the number of Monte Carlo trials, $f^{(i)}$
is the desired function value, $\hat{f}^{(i)}_j$ is the estimated value in the $j$-th Monte Carlo trial. $f_{\max}$ and 
$f_{\min}$ denote the maximum and minimum values of the function output, respectively. We also define \ac{SNR}$:=10\log(\|\bm{x}\|_2^2/\sigma_z^2)$. 

\subsection{Performance of Bit-Partitioning}

In this subsection, we evaluate the performance of the two proposed bit-partitioning methods in a network with $K=4$ nodes. The discrete inputs $x_k$ are uniformly distributed over $[0,1]$ and represented using $B =6$ bits. Moreover, we consider $L \in \{2,3,4\}$ groups and compute the \ac{NMSE} for the sum $f=\sum_{k=1}^K x_k$, the product $f=\prod_{k=1}^K x_k$, and the max $f=\max_k x_k$ functions.

Fig.~\ref{fig:uniform_bit_grouping} presents the \ac{NMSE} performance of \ac{UBP} and \ac{IABP} across a range of \ac{SNR}s. As expected, higher \ac{SNR} leads to lower \ac{NMSE} for all computed functions. Increasing $L$ spreads the available bits across more groups, which enlarges the per-group constellation spacing, thereby improving computation accuracy. In addition, for a given \ac{SNR} and $L$, \ac{IABP} achieves lower \ac{NMSE} than \ac{UBP}, since it adaptively partitions bits according to their position significance. This reduces the erroneous
reception of more significant bits and thus improves the computation accuracy, but at the cost of a higher computational complexity.




\subsection{Compare to ReMAC, SeMAC and Bit-Slicing}

\input{Figures/multi-symbol_fading_comparison}


In this subsection, we compare the proposed \ac{IABP} with other multi-symbol schemes: ReMAC, SeMAC, and Bit-Slicing. All methods are evaluated for computing the product function using $K = 4$ nodes, with input values represented by $B = 6$ bits and transmitted over $L = 2$ time slots.

As shown in Fig.~\ref{fig:comparison_L2}, \ac{IABP} achieves the lowest \ac{NMSE} across all SNR levels. Compared to \ac{SeMAC}, which maps the entire input to a sequence of modulation symbols without considering bit significance, \ac{IABP} offers better protection of more critical bits, reducing the \ac{NMSE} by approximately 5~dB. Consistent with the analysis  in~\cite{yan2025multi}, \ac{SeMAC} outperforms \ac{ReMAC} in compuatation accuracy since it explores greater diversity in modulation patterns across multiple time slots. Additionally, Bit-Slicing yields highest \ac{NMSE} because the required pre- and post-processing for nonlinear function computation introduces extra approximation errors~\cite{yan2025remac}.

\section{Conclusion}\label{sec:conclusion}

This work integrated two bit-partitioning methods into ChannelComp for multi-symbol modulation design. The uniform method \ac{UBP} maximizes the worst-case constellation distance and is addressed with a \ac{CCCP} which solves a sequence of \ac{SOCP} subproblems. The adaptive method \ac{IABP} adjusts bit-partitioning according to bit significance and updates the allocation with an outer \ac{SA} loop. Simulations show that both methods achieve reliable computation, and \ac{IABP} yields higher computation accuracy compared with \ac{UBP}. Moreover, by adapting bit allocation according to bit significance, \ac{IABP} offers stronger computation robustness than other existing multi-symbol methods. Future work will explore learning-based bit-partitioning models and extensions to scenarios of fading, imperfect channel inversion, and asynchronous transmissions.

\appendix 
\subsection{Proof of Theorem~\ref{theorem:convergence}}
\label{app:convergence}

Let $g_0(c)=-c$, $g_{i,j}(c)=c\Delta f_{i,j}$, and $h_{i,j}(\bm{x})=-\bm{x}^{\mathsf{H}}\bm{D}_{i,j}\bm{x}$, $\forall (i,j) \in [M]^2$. By convexifying the concave term in $h_{i,j}(\bm{x})$, we have $\hat{h}_{i,j}(\bm{x};\bm{x}^{(t)})=- 2\Re\{\bm{x}^{(t)\mathsf{H}}\bm{D}_{i,j}\bm{x}\} +\bm{x}^{(t)\mathsf{H}}\bm{D}_{i,j}\bm{x}^{(t)}$.

Assume $\{\bm{x}^{(t)},c^{(t)}\}$ is a feasible point for $\mathcal{P}_1$, then $\{\bm{x}^{(t)},c^{(t)}\}$ is also a feasible point for $\mathcal{P}_2$ because 
\begin{equation}
\label{eq:proof feasibility}
g_{i,j}(c^{(t)}) - \hat{h}_{i,j}(\bm{x}^{(t)};\bm{x}^{(t)}) = g_{i,j}(c^{(t)}) - {h}_{i,j}(\bm{x}^{(t)}) \leq 0.
\end{equation}
Therefore, there exists a feasible point $(\bm{x}^{(t+1)},c^{(t+1)})$ to $\mathcal{P}_2$. Given the convexity of $h_{i,j}(\bm{x})$, the following inequality holds:
\begin{equation}
\label{eq:proof convexity}
g_{i,j}(c) - {h}_{i,j}(\bm{x}) = g_{i,j}(c) - \hat{h}_{i,j}(\bm{x};\bm{x}^{(t)}) \leq 0.
\end{equation}
It then follows that $(\bm{x}^{(t+1)},c^{(t+1)})$ must be a feasible point of $\mathcal{P}_1$ since
\begin{align} 
\label{eq:proof convexity2} \nonumber
&g_{i,j}(c^{(t+1)}) - {h}_{i,j}(\bm{x}^{(t+1)}) \\
\leq & g_{i,j}(c^{(t+1)}) - \hat{h}_{i,j}(\bm{x}^{(t+1)};\bm{x}^{(t)}) \leq 0.
\end{align}
Thus, if $(\bm{x}^{(0)},c^{(0)})$ is initialized feasible, all the iterates $(\bm{x}^{(t)},c^{(t)})$ generated by Algorithm~\ref{Alg:uniform algorithm} are feasible.
In the optimization, at iteration $t$, we minimize the objective value $g_0(c)$, leading to $g_0(c^{(t)}) \geq g_0(c^{(t+1)})$. Hence, the objective sequence $\{-c^{(t)}\}_{t=0}^{\infty}$ is non-increasing and convergent.

Addtionally, based on Theorem 10 in~\cite{sriperumbudur2009convergence},
in the form of \ac{CCCP}, all the limit points of $\{\bm{x}^{(t)}\}_{t=0}^{\infty}$ are stationary points of the original problem $\mathcal{P}_0$, which concludes the proof.

\bibliographystyle{ieeetr}
\bibliography{Ref}

\end{document}

%% file: acronyms.tex

\begin{acronym}[LTE-Advanced]
  \acro{2G}{Second Generation}
  \acro{3-DAP}{3-Dimensional Assignment Problem}
  \acro{3G}{3$^\text{rd}$~Generation}
  \acro{3GPP}{3$^\text{rd}$~Generation Partnership Project}
  \acro{4G}{4$^\text{th}$~Generation}
  \acro{5G}{5$^\text{th}$~Generation}
  \acro{AA}{Antenna Array}
  \acro{AC}{Admission Control}
  \acro{AD}{Attack-Decay}
  \acro{ADC}{analog-to-digital converter}
  \acro{ADMM}{alternating direction method of multipliers}
  \acro{ADSL}{Asymmetric Digital Subscriber Line}
  \acro{AHW}{Alternate Hop-and-Wait}
  \acro{AI}{Artificial Intelligence}
  \acro{AirComp}{over-the-air computation}
  \acro{AMC}{Adaptive Modulation and Coding}
  \acro{ANN}{artificial neural network}
  \acro{AP}{\LU{A}{a}ccess \LU{P}{p}oint}
  \acro{APA}{Adaptive Power Allocation}
  \acro{ARMA}{Autoregressive Moving Average}
  \acro{ARQ}{\LU{A}{a}utomatic \LU{R}{r}epeat \LU{R}{r}equest}
  \acro{ATES}{Adaptive Throughput-based Efficiency-Satisfaction Trade-Off}
  \acro{AWGN}{additive white Gaussian noise}
  \acro{BAA}{\LU{B}{b}roadband \LU{A}{a}nalog \LU{A}{a}ggregation}
  \acro{BB}{Branch and Bound}
  \acro{BCD}{block coordinate descent}
  \acro{BD}{Block Diagonalization}
  \acro{BER}{Bit Error Rate}
  \acro{BF}{Best Fit}
  \acro{BFD}{bidirectional full duplex}
  \acro{BLER}{BLock Error Rate}
  \acro{BPC}{Binary Power Control}
  \acro{BPSK}{Binary Phase-Shift Keying}
  \acro{BRA}{Balanced Random Allocation}
  \acro{BS}{base station}
  \acro{BSUM}{block successive upper-bound minimization}
  \acro{CAP}{Combinatorial Allocation Problem}
  \acro{CAPEX}{Capital Expenditure}
  \acro{CBF}{Coordinated Beamforming}
  \acro{CBR}{Constant Bit Rate}
  \acro{CBS}{Class Based Scheduling}
  \acro{CC}{Congestion Control}
  \acro{CCCP}{constrained convex-concave procedure}
  \acro{CDF}{Cumulative Distribution Function}
  \acro{CDMA}{Code-Division Multiple Access}
  \acro{CE}{\LU{C}{c}hannel \LU{E}{e}stimation}
  \acro{CL}{Closed Loop}
  \acro{CLPC}{Closed Loop Power Control}
  \acro{CML}{centralized machine learning}
  \acro{CNR}{Channel-to-Noise Ratio}
  \acro{CNN}{\LU{C}{c}onvolutional \LU{N}{n}eural \LU{N}{n}etwork}
  \acro{CP}{computation point}
  \acro{CPA}{Cellular Protection Algorithm}
  \acro{CPICH}{Common Pilot Channel}
  \acro{CoCoA}{\LU{C}{c}ommunication efficient distributed dual \LU{C}{c}oordinate \LU{A}{a}scent}
  \acro{CoMAC}{\LU{C}{c}omputation over \LU{M}{m}ultiple-\LU{A}{a}ccess \LU{C}{c}hannels}
  \acro{CoMP}{Coordinated Multi-Point}
  \acro{CQI}{Channel Quality Indicator}
  \acro{CRM}{Constrained Rate Maximization}
	\acro{CRN}{Cognitive Radio Network}
  \acro{CS}{Coordinated Scheduling}
  \acro{CSI}{\LU{C}{c}hannel \LU{S}{s}tate \LU{I}{i}nformation}
  \acro{CSMA}{\LU{C}{c}arrier \LU{S}{s}ense \LU{M}{m}ultiple \LU{A}{a}ccess}
  \acro{CUE}{Cellular User Equipment}
  \acro{D2D}{device-to-device}
  \acro{DAC}{digital-to-analog converter}
  \acro{DC}{difference of convex}
  \acro{DCA}{Dynamic Channel Allocation}
  \acro{DE}{Differential Evolution}
  \acro{DFT}{Discrete Fourier Transform}
  \acro{DIST}{Distance}
  \acro{DL}{downlink}
  \acro{DMA}{Double Moving Average}
  \acro{DML}{Distributed Machine Learning}
  \acro{DMRS}{demodulation reference signal}
  \acro{D2DM}{D2D Mode}
  \acro{DMS}{D2D Mode Selection}
  \acro{DPC}{Dirty Paper Coding}
  \acro{DRA}{Dynamic Resource Assignment}
  \acro{DSA}{Dynamic Spectrum Access}
  \acro{DSGD}{\LU{D}{d}istributed \LU{S}{s}tochastic \LU{G}{g}radient \LU{D}{d}escent}
  \acro{DSM}{Delay-based Satisfaction Maximization}
  \acro{ECC}{Electronic Communications Committee}
  \acro{EFLC}{Error Feedback Based Load Control}
  \acro{EI}{Efficiency Indicator}
  \acro{eNB}{Evolved Node B}
  \acro{EPA}{Equal Power Allocation}
  \acro{EPC}{Evolved Packet Core}
  \acro{EPS}{Evolved Packet System}
  \acro{E-UTRAN}{Evolved Universal Terrestrial Radio Access Network}
  \acro{ES}{Exhaustive Search}
  \acro{FC}{\LU{F}{f}usion \LU{C}{c}enter}
  \acro{FD}{\LU{F}{f}ederated \LU{D}{d}istillation}
  \acro{FDD}{frequency divisionov duplex}
  \acro{FDM}{Frequency Division Multiplexing}
  \acro{FDMA}{\LU{F}{f}requency \LU{D}{d}ivision \LU{M}{m}ultiple \LU{A}{a}ccess}
  \acro{FedAvg}{\LU{F}{f}ederated \LU{A}{a}veraging}
  \acro{FER}{Frame Erasure Rate}
  \acro{FF}{Fast Fading}
  \acro{FL}{federated learning}
  \acro{FSB}{Fixed Switched Beamforming}
  \acro{FST}{Fixed SNR Target}
  \acro{FTP}{File Transfer Protocol}
  \acro{GA}{Genetic Algorithm}
  \acro{GBR}{Guaranteed Bit Rate}
  \acro{GLR}{Gain to Leakage Ratio}
  \acro{GOS}{Generated Orthogonal Sequence}
  \acro{GPL}{GNU General Public License}
  \acro{GRP}{Grouping}
  \acro{HARQ}{Hybrid Automatic Repeat Request}
  \acro{HD}{half-duplex}
  \acro{HMS}{Harmonic Mode Selection}
  \acro{HOL}{Head Of Line}
  \acro{HSDPA}{High-Speed Downlink Packet Access}
  \acro{HSPA}{High Speed Packet Access}
  \acro{HTTP}{HyperText Transfer Protocol}
  \acro{ICMP}{Internet Control Message Protocol}
  \acro{ICI}{Intercell Interference}
  \acro{ID}{Identification}
  \acro{IETF}{Internet Engineering Task Force}
  \acro{ILP}{Integer Linear Program}
  \acro{JRAPAP}{Joint RB Assignment and Power Allocation Problem}
  \acro{UID}{Unique Identification}
  \acro{IABP}{importance-adaptive bit-partitioning}
  \acro{IID}{\LU{I}{i}ndependent and \LU{I}{i}dentically \LU{D}{d}istributed}
  \acro{IIR}{Infinite Impulse Response}
  \acro{ILP}{Integer Linear Problem}
  \acro{IMT}{International Mobile Telecommunications}
  \acro{INV}{Inverted Norm-based Grouping}
  \acro{IoT}{Internet of Things}
  \acro{IP}{Integer Programming}
  \acro{IPv6}{Internet Protocol Version 6}
  \acro{ISD}{Inter-Site Distance}
  \acro{ISI}{Inter Symbol Interference}
  \acro{ITU}{International Telecommunication Union}
  \acro{JAFM}{joint assignment and fairness maximization}
  \acro{JAFMA}{joint assignment and fairness maximization algorithm}
  \acro{JOAS}{Joint Opportunistic Assignment and Scheduling}
  \acro{JOS}{Joint Opportunistic Scheduling}
  \acro{JP}{Joint Processing}
	\acro{JS}{Jump-Stay}
  \acro{KKT}{Karush-Kuhn-Tucker}
  \acro{L3}{Layer-3}
  \acro{LAC}{Link Admission Control}
  \acro{LA}{Link Adaptation}
  \acro{LC}{Load Control}
  \acro{LDC}{\LU{L}{l}earning-\LU{D}{d}riven \LU{C}{c}ommunication}
  \acro{LOS}{line of sight}
  \acro{LP}{Linear Programming}
  \acro{LSB}{least significant bit}
  \acro{LTE}{Long Term Evolution}
	\acro{LTE-A}{\ac{LTE}-Advanced}
  \acro{LTE-Advanced}{Long Term Evolution Advanced}
  \acro{LRA}{Low Rank Approximation}
  \acro{M2M}{Machine-to-Machine}
  \acro{MAC}{multiple access channel}
  \acro{MAP}{maximum a posterior}
  \acro{MANET}{Mobile Ad hoc Network}
  \acro{MC}{Modular Clock}
  \acro{MCS}{Modulation and Coding Scheme}
  \acro{MDB}{Measured Delay Based}
  \acro{MDI}{Minimum D2D Interference}
  \acro{MF}{Matched Filter}
  \acro{MG}{Maximum Gain}
  \acro{MH}{Multi-Hop}
  \acro{MIMO}{\LU{M}{m}ultiple \LU{I}{i}nput \LU{M}{m}ultiple \LU{O}{o}utput}
  \acro{MINLP}{mixed integer nonlinear programming}
  \acro{MIP}{mixed integer programming}
  \acro{MISO}{multiple input single output}
  \acro{ML}{machine learning}
  \acro{MLE}{maximum likelihood estimator}
  \acro{MLWDF}{Modified Largest Weighted Delay First}
  \acro{MME}{Mobility Management Entity}
  \acro{MMSE}{minimum mean squared error}
  \acro{MOS}{Mean Opinion Score}
  \acro{MPF}{Multicarrier Proportional Fair}
  \acro{MRA}{Maximum Rate Allocation}
  \acro{MR}{Maximum Rate}
  \acro{MRC}{Maximum Ratio Combining}
  \acro{MRT}{Maximum Ratio Transmission}
  \acro{MRUS}{Maximum Rate with User Satisfaction}
  \acro{MS}{Mode Selection}
  \acro{MSB}{most significant bit}
  \acro{MSE}{\LU{M}{m}ean \LU{S}{s}quared \LU{E}{e}rror}
  \acro{MSI}{Multi-Stream Interference}
  \acro{MTC}{Machine-Type Communication}
  \acro{MTSI}{Multimedia Telephony Services over IMS}
  \acro{MTSM}{Modified Throughput-based Satisfaction Maximization}
  \acro{MU-MIMO}{Multi-User Multiple Input Multiple Output}
  \acro{MU}{Multi-User}
  \acro{NAS}{Non-Access Stratum}
  \acro{NB}{Node B}
	\acro{NCL}{Neighbor Cell List}
  \acro{NLP}{Nonlinear Programming}
  \acro{NLOS}{non-line of sight}
  \acro{NMSE}{normalized mean square error}
  \acro{NOMA}{\LU{N}{n}on-\LU{O}{o}rthogonal \LU{M}{m}ultiple \LU{A}{a}ccess}
  \acro{NORM}{Normalized Projection-based Grouping}
  \acro{NP}{non-polynomial time}
  \acro{NRT}{Non-Real Time}
  \acro{NSPS}{National Security and Public Safety Services}
  \acro{O2I}{Outdoor to Indoor}
  \acro{OFDMA}{\LU{O}{o}rthogonal \LU{F}{f}requency \LU{D}{d}ivision \LU{M}{m}ultiple \LU{A}{a}ccess}
  \acro{OFDM}{Orthogonal Frequency Division Multiplexing}
  \acro{OFPC}{Open Loop with Fractional Path Loss Compensation}
	\acro{O2I}{Outdoor-to-Indoor}
  \acro{OL}{Open Loop}
  \acro{OLPC}{Open-Loop Power Control}
  \acro{OL-PC}{Open-Loop Power Control}
  \acro{OPEX}{Operational Expenditure}
  \acro{ORB}{Orthogonal Random Beamforming}
  \acro{JO-PF}{Joint Opportunistic Proportional Fair}
  \acro{OSI}{Open Systems Interconnection}
  \acro{PAIR}{D2D Pair Gain-based Grouping}
  \acro{PAPR}{Peak-to-Average Power Ratio}
  \acro{P2P}{Peer-to-Peer}
  \acro{PC}{Power Control}
  \acro{PCI}{Physical Cell ID}
  \acro{PDCCH}{physical downlink control channel}
  \acro{PDD}{penalty dual decomposition}
  \acro{PDF}{Probability Density Function}
  \acro{PER}{Packet Error Rate}
  \acro{PF}{Proportional Fair}
  \acro{P-GW}{Packet Data Network Gateway}
  \acro{PL}{Pathloss}
  \acro{RLT}{reformulation linearization technique}
  \acro{PRB}{Physical Resource Block}
  \acro{PROJ}{Projection-based Grouping}
  \acro{ProSe}{Proximity Services}
  \acro{PS}{\LU{P}{p}arameter \LU{S}{s}erver}
  \acro{PSO}{Particle Swarm Optimization}
  \acro{PUCCH}{physical uplink control channel}
  \acro{PZF}{Projected Zero-Forcing}
  \acro{QAM}{quadrature amplitude modulation}
  \acro{QoS}{quality of service}
  \acro{QPSK}{quadrature phase shift keying}
  \acro{QCQP}{quadratically constrained quadratic programming}
  \acro{RAISES}{Reallocation-based Assignment for Improved Spectral Efficiency and Satisfaction}
  \acro{RAN}{Radio Access Network}
  \acro{RA}{Resource Allocation}
  \acro{RAT}{Radio Access Technology}
  \acro{RATE}{Rate-based}
  \acro{RB}{resource block}
  \acro{RBG}{Resource Block Group}
  \acro{REF}{Reference Grouping}
  \acro{ReLU}{rectified linear unit}
  \acro{ReMAC}{Repetition for Multiple Access Computing}
  \acro{RF}{radio frequency}
  \acro{RLC}{Radio Link Control}
  \acro{RM}{Rate Maximization}
  \acro{RNC}{Radio Network Controller}
  \acro{RND}{Random Grouping}
  \acro{RRA}{Radio Resource Allocation}
  \acro{RRM}{\LU{R}{r}adio \LU{R}{r}esource \LU{M}{m}anagement}
  \acro{RSCP}{Received Signal Code Power}
  \acro{RSRP}{reference signal receive power}
  \acro{RSRQ}{Reference Signal Receive Quality}
  \acro{RR}{Round Robin}
  \acro{RRC}{Radio Resource Control}
  \acro{RSSI}{received signal strength indicator}
  \acro{RT}{Real Time}
  \acro{RU}{Resource Unit}
  \acro{RUNE}{RUdimentary Network Emulator}
  \acro{RV}{Random Variable}
  \acro{SA}{simulated annealing}
  \acro{SAA}{Small Argument Approximation}
  \acro{SAC}{Session Admission Control}
  \acro{SCM}{Spatial Channel Model}
  \acro{SC-FDMA}{Single Carrier - Frequency Division Multiple Access}
  \acro{SD}{Soft Dropping}
  \acro{S-D}{Source-Destination}
  \acro{SDPC}{Soft Dropping Power Control}
  \acro{SDMA}{Space-Division Multiple Access}
  \acro{SDR}{semidefinite relaxation}
  \acro{SDP}{semidefinite programming}
  \acro{SeMAC}{Sequential Modulation for AirComp}
  \acro{SeMAC-PA}{SeMAC with Power Adaptation}
  \acro{SER}{Symbol Error Rate}
  \acro{SES}{Simple Exponential Smoothing}
  \acro{S-GW}{Serving Gateway}
  \acro{SGD}{\LU{S}{s}tochastic \LU{G}{g}radient \LU{D}{d}escent}  
  \acro{SINR}{signal-to-interference-plus-noise ratio}
  \acro{SI}{self-interference}
  \acro{SIP}{Session Initiation Protocol}
  \acro{SISO}{\LU{S}{s}ingle \LU{I}{i}nput \LU{S}{s}ingle \LU{O}{o}utput}
  \acro{SIMO}{Single Input Multiple Output}
  \acro{SIR}{Signal to Interference Ratio}
  \acro{SLNR}{Signal-to-Leakage-plus-Noise Ratio}
  \acro{SMA}{Simple Moving Average}
  \acro{SNR}{\LU{S}{s}ignal-to-\LU{N}{n}oise \LU{R}{r}atio}
  \acro{SOCP}{second-order cone programming}
  \acro{SORA}{Satisfaction Oriented Resource Allocation}
  \acro{SORA-NRT}{Satisfaction-Oriented Resource Allocation for Non-Real Time Services}
  \acro{SORA-RT}{Satisfaction-Oriented Resource Allocation for Real Time Services}
  \acro{SPF}{Single-Carrier Proportional Fair}
  \acro{SRA}{Sequential Removal Algorithm}
  \acro{SRS}{sounding reference signal}
  \acro{SU-MIMO}{Single-User Multiple Input Multiple Output}
  \acro{SU}{Single-User}
  \acro{SVD}{singular value decomposition}
  \acro{SVM}{\LU{S}{s}upport \LU{V}{v}ector \LU{M}{m}achine}
  \acro{TCP}{Transmission Control Protocol}
  \acro{TDD}{time division duplex}
  \acro{TDMA}{\LU{T}{t}ime \LU{D}{d}ivision \LU{M}{m}ultiple \LU{A}{a}ccess}
  \acro{TNFD}{three node full duplex}
  \acro{TETRA}{Terrestrial Trunked Radio}
  \acro{TP}{Transmit Power}
  \acro{TPC}{Transmit Power Control}
  \acro{TTI}{transmission time interval}
  \acro{TTR}{Time-To-Rendezvous}
  \acro{TSM}{Throughput-based Satisfaction Maximization}
  \acro{TU}{Typical Urban}
  \acro{UBP}{uniform bit-partitioning}
  \acro{UE}{\LU{U}{u}ser \LU{E}{e}quipment}
  \acro{UEPS}{Urgency and Efficiency-based Packet Scheduling}
  \acro{UL}{uplink}
  \acro{UMTS}{Universal Mobile Telecommunications System}
  \acro{URI}{Uniform Resource Identifier}
  \acro{URM}{Unconstrained Rate Maximization}
  \acro{VR}{Virtual Resource}
  \acro{VoIP}{Voice over IP}
  \acro{WAN}{Wireless Access Network}
  \acro{WCDMA}{Wideband Code Division Multiple Access}
  \acro{WF}{Water-filling}
  \acro{WiMAX}{Worldwide Interoperability for Microwave Access}
  \acro{WINNER}{Wireless World Initiative New Radio}
  \acro{WLAN}{Wireless Local Area Network}
  \acro{WMMSE}{weighted minimum mean square error}
  \acro{WMPF}{Weighted Multicarrier Proportional Fair}
  \acro{WPF}{Weighted Proportional Fair}
  \acro{WSN}{Wireless Sensor Network}
  \acro{WWW}{World Wide Web}
  \acro{XIXO}{(Single or Multiple) Input (Single or Multiple) Output}
  \acro{ZF}{Zero-Forcing}
  \acro{ZMCSCG}{Zero Mean Circularly Symmetric Complex Gaussian}
\end{acronym}

%% file: Figures/uniform_Alg.tex


\begin{algorithm}[!t]\label{Alg:uniform algorithm}
\caption{\ac{CCCP} for \ac{UBP}}

\textbf{Given} an initial point $\bm{x}^{(0)}$. Set $t=0$.

\textbf{Repeat}

    \Indp
    {\it Convexify}. Linearize $-\bm{x}^{\mathsf{H}}\bm{D}_{i,j}\bm{x}$ to obtain $\mathcal{P}_2$. \\

    {\it Solve}. Obtain $\bm{x}^{(t+1)}$ by solving $\mathcal{P}_{2}$. \\

    
    {\it Update iteration}. $t:=t+1$.
    
    \Indm
    
    \textbf{Until} $c^{(t+1)}-c^{(t)}<\delta$ is satisfied.

\textbf{Output} ${\hat{\bm{x}}} \gets \bm{x}^{(t)}$.
 
\end{algorithm}

%% file: Figures/importance_aware_Alg.tex
\begin{algorithm}[!t]\label{Alg:importance-aware algorithm}
\caption{\ac{CCCP} and \ac{SA} for \ac{IABP}}

\textbf{Given} an initial vector $\bm{b}^{(0)}$, $\phi_0>0$, $\phi_{\min}$, $\alpha \in (0,1)$. Set $r=0$.

\textbf{Repeat}

    \Indp
    \textbf{Given} an initial point $\{\bm{x}_\ell^{(0)}\}$, candidate  vector $\bm{b}'$. Set $t=0$. \\
    
    \textbf{Repeat}

        \Indp
        {\it Convexify}. Linearize $-\bm{x}_\ell^{\mathsf{H}}\bm{D}_{i,j,\ell}\bm{x}_\ell$ to obtain $\mathcal{P}_5$ \\
    
        {\it Solve}. Obtain $\{\bm{x}^{(t+1)}_\ell\}$ by solving $\mathcal{P}_{5}$. \\
    
        
        {\it Update iteration}. $t:=t+1$.
        
        \Indm
    
    \textbf{Until} $c^{(t+1)}-c^{(t)} < \delta$ is satisfied.

    {\it Update $b$}. Accept $\bm{b}^{(r+1)}=\bm{b}'$ with probability $p$. \\ $\qquad$ $\qquad$ Otherwise, $\bm{b}^{(r+1)}=\bm{b}^{(r)}$.

    {\it Update $\phi$}. $\phi_{r+1}=\alpha \phi_r$. \\
        
    {\it Update iteration}. $r:=r+1$.
    
    \Indm

\textbf{Until} $\phi_r<\phi_{\min}$ is satisfied.    

\textbf{Output} $\hat {\bm{b}} \gets \bm{b}^{(r)}$, $\{{\hat{\bm{x}}_\ell}\} \gets \{\bm{x}_\ell^{(t)}\}$.
 
\end{algorithm}
\vspace{-1em}

%% file: Figures/uniform_bit_grouping.tex
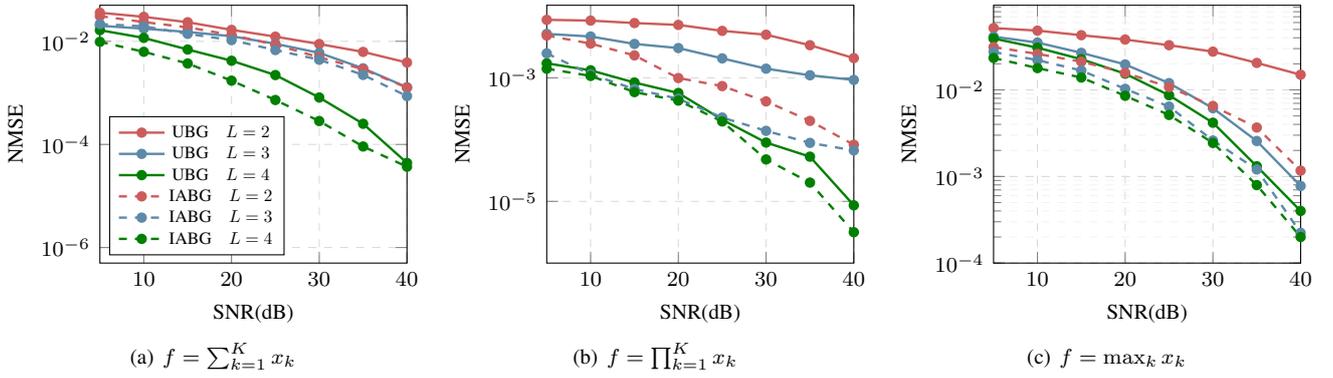
\begin{figure*}[!t]
\centering
\subfigure[$f = \sum_{k=1}^Kx_k$]{   \label{fig:uniform_four_node_sum_SNR}
\begin{tikzpicture}
    \begin{axis}[
        xlabel = {SNR(dB)},
        ylabel = {NMSE},
        label style={font=\footnotesize},
        width=0.31\textwidth,
        height=5cm,
        xmin=5, xmax=40,
        ymin=0.0000005, ymax=0.05,
        legend style={nodes={scale=0.65, transform shape}, at={(0.3,0.85)}},
        ticklabel style = {font=\footnotesize},
        legend pos=south west,
        ymajorgrids=true,
        xmajorgrids=true,
        grid style=dashed,
        grid=both,
        ymode = log,
        grid style={line width=.1pt, draw=gray!10},
        major grid style={line width=.2pt,draw=gray!30},
    ]
    \addplot[
        color=chestnut,
        mark=*,
        line width=0.9pt,
        mark size=1.5pt,
        ]
    table[x=SNR,y=sum6_L2]
    {Data/Sim_SNR_CCP.dat};
    \addplot[ 
            color=airforceblue,
            mark=*,
            line width=0.9pt,
            mark size=1.5pt,
            ]
    table[x=SNR,y=sum6_L3]
    {Data/Sim_SNR_CCP.dat};
    \addplot[ 
        color=cssgreen,
        mark=*,
        line width=0.9pt,
        mark size=1.5pt,
        ]
    table[x=SNR,y=sum6_L4]
    {Data/Sim_SNR_CCP.dat};
    \addplot[
             dashed,
        color=chestnut,
        mark=*,
        mark options = {rotate = 180, solid},
        line width=0.9pt,
        mark size=1.5pt,
        ]
    table[x=SNR,y=sum6_I2]
    {Data/Sim_SNR_CCP.dat};
    \addplot[ 
              dashed,
            color=airforceblue,
            mark=*,
            mark options = {rotate = 180, solid},
            line width=0.9pt,
            mark size=1.5pt,
            ]
    table[x=SNR,y=sum6_I3]
    {Data/Sim_SNR_CCP.dat};
    \addplot[ 
             dashed,
        color=cssgreen,
        mark=*,
        mark options = {rotate = 180, solid},
        line width=0.9pt,
        mark size=1.5pt,
        ]
    table[x=SNR,y=sum6_I4]
    {Data/Sim_SNR_CCP.dat};
    \legend{UBG$\quad L=2$,UBG$\quad L=3$,UBG$\quad L=4$,IABG$\quad L=2$,IABG$\quad L=3$,IABG$\quad L=4$};
\end{axis}
\end{tikzpicture}
}\subfigure[$f=\prod_{k=1}^Kx_k$]{\label{fig:uniform_four_node_prod_SNR}
\begin{tikzpicture}
    \begin{axis}[
        xlabel = {SNR(dB)},
        ylabel = {NMSE},
        label style={font=\footnotesize},
        width=0.31\textwidth,
        height=5cm,
        xmin=5, xmax=40,
        ymin=0.000001, ymax=0.015,
        legend style={nodes={scale=0.65, transform shape}, at={(0.3,0.85)}},
        ticklabel style = {font=\footnotesize},
        legend pos=south west,
        ymajorgrids=true,
        xmajorgrids=true,
        grid style=dashed,
        grid=both,
        ymode = log,
        grid style={line width=.1pt, draw=gray!10},
        major grid style={line width=.2pt,draw=gray!30},
    ]
    \addplot[
        color=chestnut,
        mark=*,
        line width=0.9pt,
        mark size=1.5pt,
        ]
    table[x=SNR,y=prod6_L2]
    {Data/Sim_SNR_prod_CCP.dat};
    \addplot[ 
            color=airforceblue,
            mark=*,
            line width=0.9pt,
            mark size=1.5pt,
            ]
    table[x=SNR,y=prod6_L3]
    {Data/Sim_SNR_prod_CCP.dat};
    \addplot[ 
        color=cssgreen,
        mark=*,
        line width=0.9pt,
        mark size=1.5pt,
        ]
    table[x=SNR,y=prod6_L4]
    {Data/Sim_SNR_prod_CCP.dat};
    \addplot[
             dashed,
        color=chestnut,
        mark=*,
        mark options = {rotate = 180, solid},
        line width=0.9pt,
        mark size=1.5pt,
        ]
    table[x=SNR,y=prod6_I2]
    {Data/Sim_SNR_prod_CCP.dat};
    \addplot[ 
              dashed,
            color=airforceblue,
            mark=*,
            mark options = {rotate = 180, solid},
            line width=0.9pt,
            mark size=1.5pt,
            ]
    table[x=SNR,y=prod6_I3]
    {Data/Sim_SNR_prod_CCP.dat};
    \addplot[ 
             dashed,
        color=cssgreen,
        mark=*,
        mark options = {rotate = 180, solid},
        line width=0.9pt,
        mark size=1.5pt,
        ]
    table[x=SNR,y=prod6_I4]
    {Data/Sim_SNR_prod_CCP.dat};
\end{axis}
\end{tikzpicture}}
\subfigure[$f=\max_{k}x_k$]{\label{fig:uniform_four_node_max_SNR}
\begin{tikzpicture}
    \begin{axis}[
        xlabel = {SNR(dB)},
        ylabel = {NMSE},
        label style={font=\footnotesize},
        width=0.31\textwidth,
        height=5cm,
        xmin=5, xmax=40,
        ymin=0.0001, ymax=0.095,
        legend style={nodes={scale=0.65, transform shape}, at={(0.3,0.85)}},
        ticklabel style = {font=\footnotesize},
        legend pos=south west,
        ymajorgrids=true,
        xmajorgrids=true,
        grid style=dashed,
        grid=both,
        ymode = log,
        grid style={line width=.1pt, draw=gray!10},
        major grid style={line width=.2pt,draw=gray!30},
    ]
    \addplot[
        color=chestnut,
        mark=*,
        line width=0.9pt,
        mark size=1.5pt,
        ]
    table[x=SNR,y=max6_L2]
    {Data/Sim_SNR_max_CCP.dat};
    \addplot[ 
            color=airforceblue,
            mark=*,
            line width=0.9pt,
            mark size=1.5pt,
            ]
    table[x=SNR,y=max6_L3]
    {Data/Sim_SNR_max_CCP.dat};
    \addplot[ 
        color=cssgreen,
        mark=*,
        line width=0.9pt,
        mark size=1.5pt,
        ]
    table[x=SNR,y=max6_L4]
    {Data/Sim_SNR_max_CCP.dat};
    \addplot[
             dashed,
        color=chestnut,
        mark=*,
        mark options = {rotate = 180, solid},
        line width=0.9pt,
        mark size=1.5pt,
        ]
    table[x=SNR,y=max6_I2]
    {Data/Sim_SNR_max_CCP.dat};
    \addplot[ 
              dashed,
            color=airforceblue,
            mark=*,
            mark options = {rotate = 180, solid},
            line width=0.9pt,
            mark size=1.5pt,
            ]
    table[x=SNR,y=max6_I3]
    {Data/Sim_SNR_max_CCP.dat};
    \addplot[ 
             dashed,
        color=cssgreen,
        mark=*,
        mark options = {rotate = 180, solid},
        line width=0.9pt,
        mark size=1.5pt,
        ]
    table[x=SNR,y=max6_I4]
    {Data/Sim_SNR_max_CCP.dat};
\end{axis}
\end{tikzpicture}
}
\caption{Performance of the proposed \ac{UBP} under different SNRs in terms of NMSE averaged over $N_s=100$. The number of bits are $B =6$ and the desired functions are $f=\sum_{k=1}^K x_k$, $f=\prod_{k=1}^K x_k$ and $f=\max_k x_k$ with $K=4$ nodes.}
\label{fig:uniform_bit_grouping}
\end{figure*}

%% file: Figures/multi-symbol_fading_comparison.tex
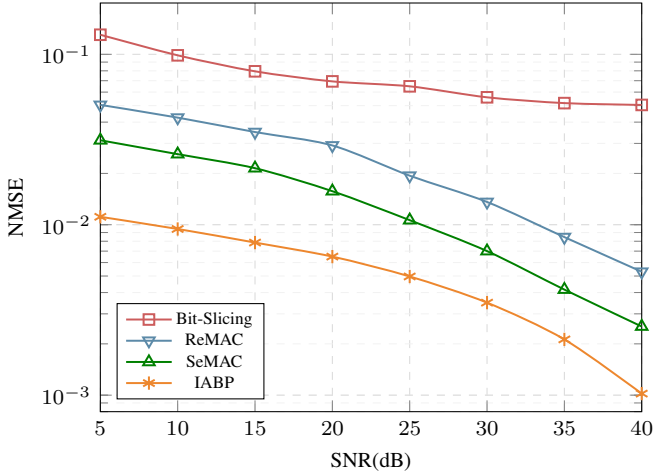
\begin{figure}[!t]
\centering
\begin{tikzpicture}
    \begin{axis}[
        xlabel = {SNR(dB)},
        ylabel = {NMSE},
        label style={font=\footnotesize},
        width=0.48\textwidth,
        height=7cm,
        xmin=5, xmax=40,
        ymin=0.0008, ymax=0.2,
        legend style={nodes={scale=0.65, transform shape}, at={(0.3,0.85)}},
        ticklabel style = {font=\footnotesize},
        legend pos=south west,
        ymajorgrids=true,
        xmajorgrids=true,
        grid style=dashed,
        grid=both,
        ymode = log,
        grid style={line width=.1pt, draw=gray!10},
        major grid style={line width=.2pt,draw=gray!30},
    ]
    \addplot[ smooth,
             thin,
        color=chestnut,
        mark=square,
        line width=0.75pt,
        mark size=2pt,
        ]
    table[x=SNR,y=bs_prod]
    {Data/Sim_comparison.dat};
    \addplot[ smooth,
             thin,
        color=airforceblue,
        mark=triangle,
        mark options = {rotate = 180, solid},
        line width=0.75pt,
        mark size=2.5pt,
        ]
    table[x=SNR,y=ReMAC_prod]
    {Data/Sim_comparison.dat};
    \addplot[ smooth,
             thin,
        color=cssgreen,
        mark=triangle,
        line width=0.75pt,
        mark size=2.5pt,
        ]
    table[x=SNR,y=SeMAC_prod]
    {Data/Sim_comparison.dat};
    \addplot[ smooth,
             thin,
        color=cadmiumorange,
        mark=asterisk,
        mark options = {rotate = 180, solid},
        line width=0.75pt,
        mark size=2.5pt,
        ]
    table[x=SNR,y=bg_prod_CCP]
    {Data/Sim_comparison.dat};
\legend{Bit-Slicing ,ReMAC ,SeMAC, IABP};
\end{axis}
\end{tikzpicture}
\vspace{-1em}
\caption{Performance comparison among ReMAC, \ac{SeMAC}, Bit-Slicing and \ac{IABP} with \( K = 4 \) nodes 
across $L=2$ time slots for the product function. The input value $x_k$ is quantized by $B=6$ bits.
}
\label{fig:comparison_L2}
\end{figure}